\documentclass[acus]{JAC2000}

\usepackage{graphicx}

\begin{document}
\title{\flushright{FRBT005}\\[15pt] \centering SIGNAL ARCHIVING AND RETRIEVAL: 
ESSENTIAL LONG TERM PERFORMANCE TUNING TOOL\thanks{
Funded by the Bundesministerium f\"ur Bildung, Wissenschaft, Forschung und
Technologie (BMBF) and the Land Berlin}}

\author{ R. Bakker, T. Birke, R. M\"uller, BESSY, Berlin, Germany}

\maketitle

\begin{abstract}
The first two years of user service of the third generation light source
BESSY II emphasized the importance of a reliable, comprehensive and dense
logging of a few thousand setpoints, readbacks, status and alarm values.
Today data from sources with various characteristics residing in different
protected networks are centrally collected and retrievable via an uncomplex
CGI program to any desktop system on the site.  Data post-processing tools
cover Windows applications, IDL, SDDS and custom programs matching users
skills and preferences.  In this paper illustrative sample data
explorations are described that underline the importance of the logging
system for operations as well as for the understanding of singular events
or long term drifts. Serious shortcomings of the present installation and
focus of further development are described.
\end{abstract}

\section{Introduction}

Like other third generation light sources BESSY II exceeds many of the
primary design goals. Users not only appreciate the additional potential of
the excellent beam definition and stability --- an increasing number of
experiments simply depend on the high and reliable beam quality. Especially
important are minimal beam center of mass drifts, well defined beam energy
with minimal spread and high beam intensity with a long lifetime. It is
difficult to prevent drifting of these parameters over the several days
necessary for an experiment. Many effects originating from facility
operating conditions and user activities can contribute. Since there is
never enough time to isolate all possible effects by dedicated accelerator
development studies archives of logged data are most important sources of
information.

\section{Setup and Status}

Despite the eminent importance of archived data the archiving system at
BESSY is far from being well settled. Adequate carefully done is the
collection both of snapshot files and long term monitoring data
\cite{ica99}. Loss or omission of essential data would destroy
unrecoverable knowledge about past behaviour of the facility.  Retrieval
tools are still cumbersome, immature and subject of maloperation frequently
resulting in loss of work time. Configuration is mainly hand-work, thus not
fault free. Surveillance of data source availability and data integrity is
done occasionally. Only the collector programs themselves are
systematically supervised by watch-dog or stop/restart procedures.

\subsection{SDDS based Data Store}
Initially the BESSY archiving configuration was based on the SDDS toolkit.
Storage format are compressed SDDS files spanning a device class and a full
day, sorted into a calendar mapping directory structure.  A TclTk glue
application combines navigation, SDDS data retrieval, correlation and
export~\cite{ica99}.

This data store is still a good compromise even though not optimal with
respect to data format and size, network resources and CPU requirements:
channel selection, previewing facility, available post-processing tools
cover most of operators requirements.  The SDDS archive is not
discontinued, collects 20 GB/y and serves as valuable backup system.  A
more or less frozen and easy maintainable list of signals essential for the
understanding of basic operation parameters are monitored. Major obstacle
for a site-wide usage of the archive is the (intended) in-accessibility of
the data store residing in the protected accelerator control production
area.

\subsection{Central Channel Archiver}
\label{CA}
Since mid 2000 a {\em Channel Archiver}~\cite{Kay} instance has been set up
in addition. It is intended to overcome the self-containment of the
(accelerator) SDDS archiver and serve the whole site. Any major development
and configuration effort goes into this system. Data collector engine(s)
and {\tt CGIExport} retrieval tools are installed in a dedicated
environment\cite{ica99}. A six processor HP N-class server ({\em archive
server}) in a non-routable private network stores the data on a RAID system
that is backed up to a tape robot. It is planned to migrate mass storage to
a fibre channel system attached to a tape library this year.

\subsection{Data Flow}
In an attempt to minimize adverse effects on the system caused by
unexpected activities and to maximize uptime neither user accounts nor NFS
access to the archiver network are provided.  For data collection all data
sources residing on dedicated networks are connected by two multi-homed
CA-gateway computers (8 network interfaces each).  Presently a single
archiving engine (process) stores 50 GB/y accelerator relevant data. A
second engine has been set up early this year for the beamline area and
auxiliary data presently collecting about 15GB/y.

Common retrieval method is {\tt HTTP} invocation of {\tt CGIExport}
\cite{Kay} via the central network router. Typically the available {\tt gnuplot}
presentation of the data requested is used as a preview ensuring that the
data selection provides the desired information. Then the data are
retrieved in spread-sheet or matlab format and stored on local
disk. Favourite postprocessing tools are PC Windows tools (Origin, Excel)
or UNIX applications (IDL, Matlab). A small program ({\tt caa2sdds})
converts the spread-sheet output to SDDS format enabling data analysis with
the full data selection, post-processing and display power of SDDS.

%For an improved comfort of data identification and previewing it is
%presently considered to allow ssh access to the archive server utilizing
%the public key authentification: This allows `known' users to start up and
%utilize the X-browser {\tt Xarr}. Especially interactive zoom and clip
%capabilities of this archive client help to find regions of interest.

\section{Typical Utilization}

\subsection{Identification of Singular Events}
Probably tracking down sudden perturbations to its causes is the most
common usage of the archive. Examples for this application are e.g.\ an unusual
large drift that corresponded to the failure of a water pump or the sudden onset
of orbit jumps that was due to an improper motor reset resulting in a constant
rotation of strong chicane magnets.

\subsection{First Hints on Unexpected Effects}

\begin{figure}[ht]
\centering
\includegraphics*[height=80mm,angle=270]{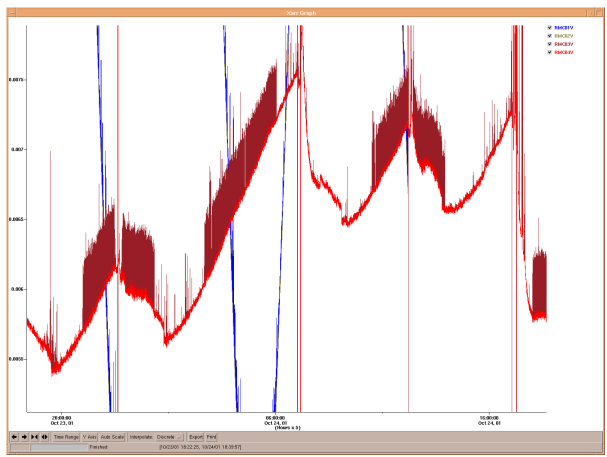}
\caption{Orbit perturbations due to switched user magnet outside the
  storage ring tunnel. Phases of experimental activity are clearly
  visible. Viewing tool: {\tt Xarr}.}
\label{User}
\end{figure}

Archived data help to get a first idea of possible
explanations: Mid 2001 for example a strong, periodic orbit
perturbation has been reported by the operators. By phase analysis it was
possible to locate the problem source with a few meters precision
at a ring segment where no active elements are installed.  The time pattern
of perturbation onset and disappearance (see fig.\ \ref{User}) suggested an
unknown correlation with user activities. Targeted investigation found out
that one user group reversed the field of a 1 [T] magnet twice a minute 
several meters apart from the beampipe.

%% ---------------------------------------------------------------------
%% ---------------------------------------------------------------------

\subsection{Analysis of Changes}
%% ---------------------------------------------------------------------
\begin{figure}[hbt]
\centering
\includegraphics*[width=80mm,angle=0]{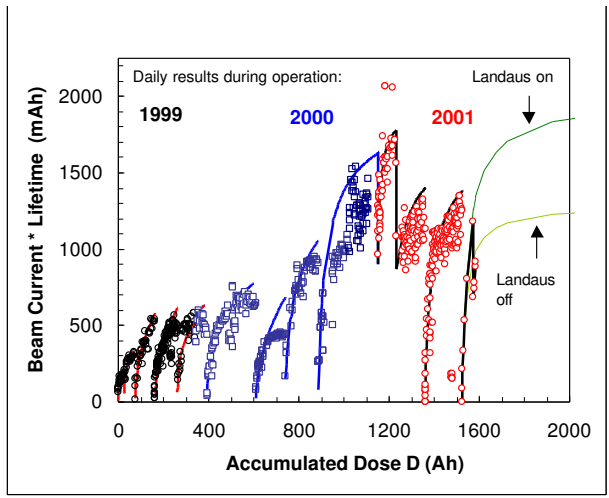}
\caption{Vacuum effects on lifetime shown as a function of accumulated 
beam dose. Postprocessing  tool: {\tt EXCEL}}
\label{Dose}
\end{figure}
%% ---------------------------------------------------------------------

On the long term extreme the archive provides the data needed to make
evolutions visible that are hardly perceptible on a fill to fill basis.
Plotting e.g. the normalized lifetime [mAh] against the accumulated dose
[Ah] over the full operating time of the facility is a powerful mean to
find out very fundamental factors: From fig.\ \ref{Dose} it can be
concluded, that the vacuum related lifetime reduction is basically overcome
by beam scrubbing 1000 [Ah] after start up of the accelerator. Every
venting due to installation requirements needs another 100 [Ah] to
reinstall the previous performance. On top of these basic conditions global
lifetime improving effects of Landau cavities (mid 2000) as well as
reducing effects of imperfectly corrected insertion devices (beginning
2001) can be seen.

\section{Demanding Requirements}

\subsection{Uptime, Reliability}
Requirements on uptime, reliability and consistency of the archive are
substantial.  The archive data have to contain signals of very different
importance.  Beam intensity e.g.\ is analyzed and correlated in any
thinkable way: integration (dose), differentiation (beam loss), pattern
analysis (user runs) etc. Here a loss of data would be serious, but
recognized within minutes.  Other signals are monitored as a
precaution. They could potentially help to find candidates for sources of
performance degradation. Dispensable for the all day business they are not
under human surveillance. Regardless they have to contain reliable data
when needed.

\subsection{Data Density, Aging}

%% ---------------------------------------------------------------------
\begin{figure}[ht]
\centering
\includegraphics*[width=80mm,angle=0]{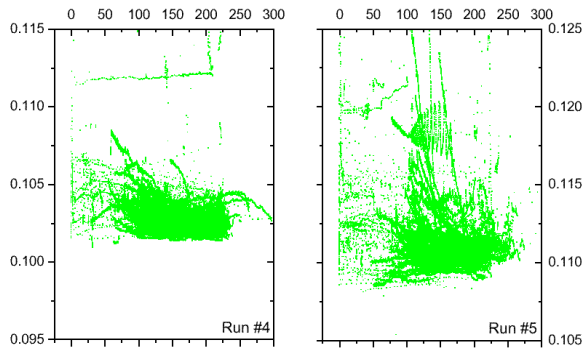}
\caption{Raw data of (uncorrected) vertical orbit stability (+/- 5 $\mu$m
  RMS) during all user fills (220 mA - 80 mA) at user run \#4
  (Aug. 2000, left). General degradation and spurious exotic drifts can be clearly
  identified at run \#5 (Oct. 2000, right). Postprocessing Tool: {\tt Origen}}
\label{Orbit}
\end{figure}
%% ---------------------------------------------------------------------

\begin{figure}[ht]
\centering
\includegraphics*[width=65mm,angle=0]{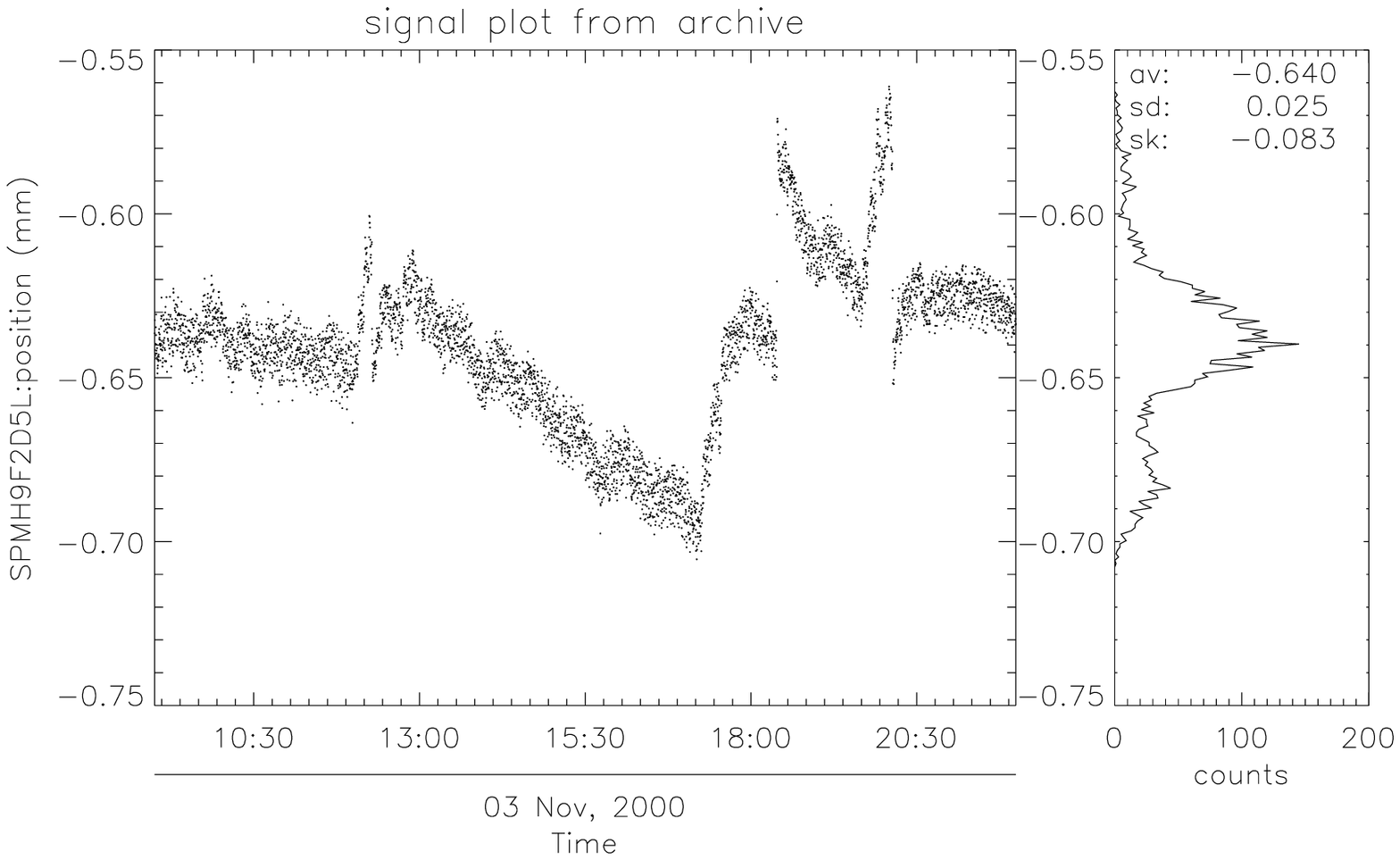}
\includegraphics*[width=65mm,angle=0]{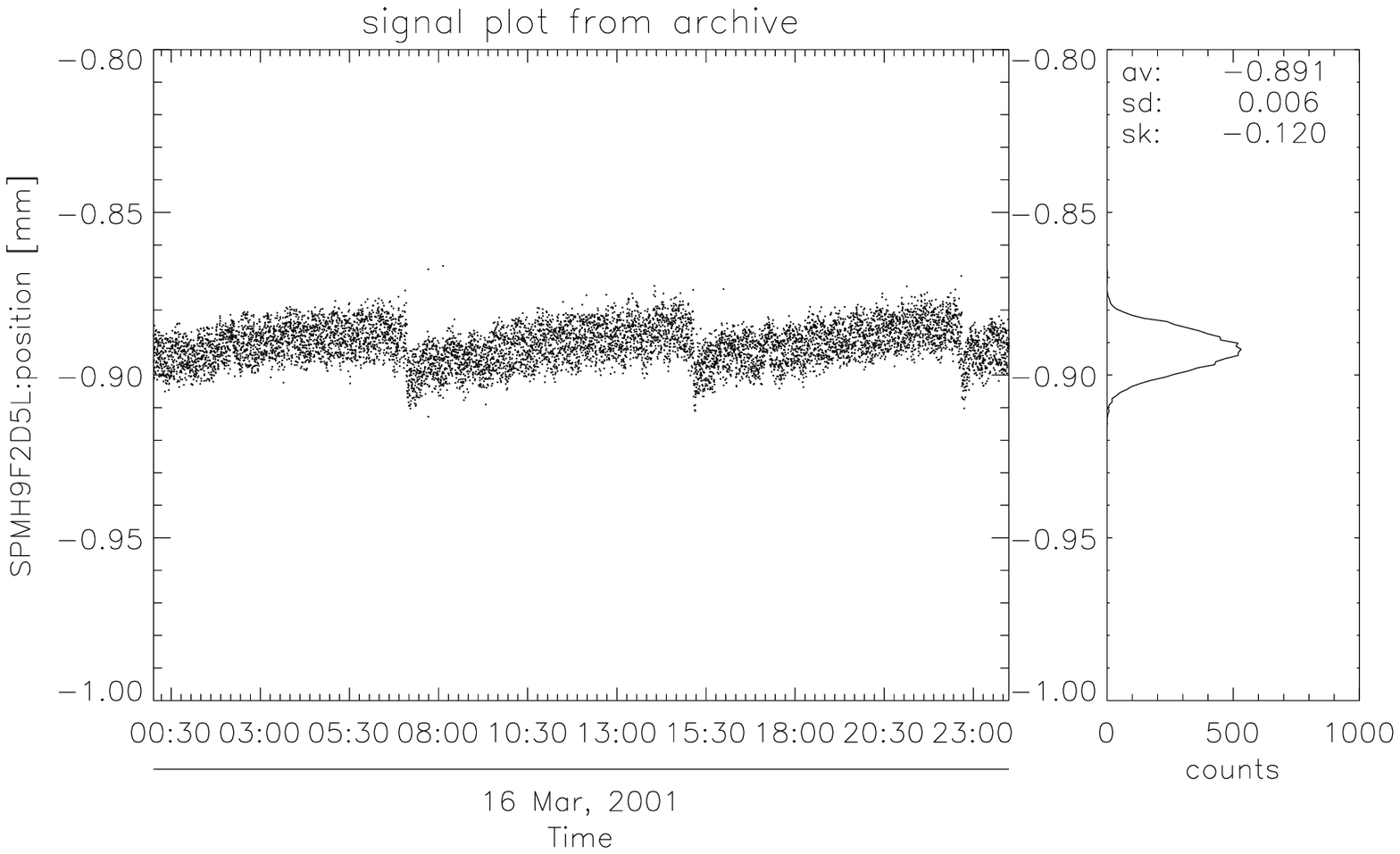}
\caption{Stability comparison of `Uncorrected'  with `Drift Corrected'
  Fills. Data  Postprocessing Tool: {\tt IDL}.}
\label{Fill}
\end{figure}

The most common approaches to prevent growing of the archive to unmanageable
dimensions are removal of `old' data (tape, deletion) or a progressive
reduction of data density. Fig. \ref{Orbit} and \ref{Fill} are examples of
the opposite requirement for a dense and long term archive. In
Fig. \ref{Orbit} spurious observations and user complaints could be
quantified after serious hardware modifications. The comparison of
performance and influence of a new operation mode required per fill details
(8h) months apart for fig.\ \ref{Fill}.

\section{Present Focus of Activities}
\subsection{Data Collector}
Today management and configuration of collector engines is further
robustified.  Usage of the system is simplified by GUI administration
tools. Signal configuration management based on the reference RDB is still
missing.

\subsection{Retrieval}

Performance of data retrieval from large and multiple archives has been
drastically enhanced. Channel detection method for a given time interval
is improved.  Volume of intermediate data needed for previewing is reduced
to the minimum allowed by the anticipated {\tt gnuplot} resolution.

%% ---------------------------------------------------------------------
%% ---------------------------------------------------------------------

\subsection{Data Partitioning}

From the iterator model and the hash table directories the binary data
format of the {\em Channel Archiver} is optimized for retrieval of data
from archives containing a moderate number of channels and starting
e.g.\ from `now' going backwards in time. Retrieving a dozen of channels out
of the `middle' of a continuous archive holding several thousands of
channels requires patience.

As a first improvement approach the huge monolithic data
block is split into a moderate number of weekly ordered chunks holding certain
fragments of the whole signal collection. Adjustment of the I/O routines
results in orders of magnitude retrieval acceleration. But however
home grown data formats are optimized: ultimately the retrieval of
arbitrary data selections out of huge data stores is best done with
commercial RDB systems.  Consequently the utilization of a RDB storage
format has to be re-considered.

\section{SUMMARY}

Ideally one would like to be able to `replay' any controllable and
measurable parameter out of the signal archive  with the reasonable time
resolution of a few seconds. For a BESSY size
facility this would require data stores of several TB/y.  The {\em Channel
Archiver} provides a robust data collector and retrieval toolkit but the
archive itself has to be reduced to manageable dimensions.

The challenges today are configuration (select relevant signals, grouping, choose
proper archiving frequencies), correlation detection (identify signals) and
data organisation (optimized search). Plotting options and postprocessing
requirements have to be provided by the end-user according to his specific
skills and varying needs.

\end{document}